\documentclass[preprint,aps]{revtex4-1}
\usepackage{graphicx}
\usepackage{rotating}
\usepackage{amsmath}
\usepackage{bm}

\begin{document}
\title{ Low temperature Universalities in amorphous systems: role of microscopic length scales}
\author{Pragya Shukla}
\affiliation{Department of Physics, Indian Institute of Technology, Kharagpur-721302, W.B., India  }
\date{\today}

\widetext

\begin{abstract}

We find that  a competition between dispersion forces among  molecules in solids and their phonon mediated coupling  leads to  a natural length scale  based on molecular parameters and relevant to decipher glass anomalies. For amorphous systems, the length scale  is of the medium range topological orders and its ratio  with the distance of closest approach between molecules turns out to be a constant.  This in turn leads to a material independent, constant value of the  ratio ${\gamma_l \over \gamma_t}$, with $\gamma_l$ and $\gamma_t$ as the  coupling-strength for two amorphous molecules mediated by longitudinal and transverse phonons (also referred as Meissner-Berret ratio) and thereby provides a theoretical explanation of their experimentally obserevd quantitative universality in \cite{mb}. The above length scales are also related to Ioffe-Regel frequency  and boson peak frequency  of the vibrational spectrum and indicate that the former is of the same order as the latter for transverse phonon-dynamics.

\end{abstract}

\maketitle
.


\section{Introduction}

Experimental observation of anamalous but universal thermal and acoustic properties in a wide range of 
disordered solids  at low temperatures  has motivated an intense quest for a theoretical explanation \cite{plt}. The behaviour, often referred as the ''glassy behavior'', has been attributed to presence of certain defects in the solid which couple with phonon fields and lead to deviation of the properties from crystals.   To justify the universality, an important requirement expected to be fulfilled by the theoretical explanation is that is should be based on  an unified form of the defects among amorphous solids.
The search for such defects has motivated many models in past, with one of them, namely, the two level tunnelling systems (TTLS) achieving remarkable success \cite{and,phil}.  The other important suggestions in this context are the soft potential model (including quasi-localized vibtations along with TLS as defects) \cite{spm,paras,buch}, fractons \cite{frac,gg1}, harmonic oscillator models with disorder in force constants \cite{sdg}, models with random spatial variations in elastic moduli \cite{schi},  effective medium theories (e.g.\cite{emt, degi}), Euclidean random matrices \cite{grig}, characteristic vibrations of clusters of medium range order (MRO) \cite{du,vdos,ell3,sksq}, renormalisation approach \cite{vl,dl} etc.  Based on recent experiments \cite{mizu, lern}, it is now generally believed that low temperature universalities are a reflection of medium range order peculiarities in glasses \cite{du,vdos,ell3,sksq,mg,degi}. This point of view is also supported  by our recent study of specific heat of a MRO size sample \cite{bb1} and internal friction of a macroscopic sample of amorphous material \cite{qc1}. The primary focus of the present work is provide further theoretical support in this context.

Based on experimental studies, the topological order in the solid  can be divided into three parts: long range, medium range and short range. In the non-crystalline phase, the long range order (LRO) is absent, the medium range order (MRO) is over distances of $20-30 \AA$ and the short range order (SRO) refers to first and 2nd nearest neighbor distances. The bonding interactions at short range are dictated by the valance force fields (covalent or ionic, also acting as mechanical constraints)  but the medium range order is governed by the Vanderwaal (VW) interactions. The relative weakness of the latter (compared to covalent bonding) leads to considerable flexibility in the local structure and the first appearance of randomness (in comparison to crystals) \cite{blhf,ell}. For example, for  $SiO_2$, the simplest structural unit in the 3D case is a network of corner-sharing $SiO_4$ tetrahedra, interconnected by VW interactions. In $B_2O_3$, medium range order contains linked triangle forming planar rings, referred as Boroxyl rings, with strong intra-ring bonds and VW inter-rings bonds. For distances beyond $30 \AA$ typically, the VW interactions become very weak  and other types of interactions begin to dominate the structure. In crystalline material, this leads to a long range order (LRO)  but the latter is absent in the non-crystalline phase. In metallic glasses, this range has also been shown to have power-law scaling and fractal nature. In some glasses e.g. $a-Si$, the medium range order is found to extend even to $35 A$ \cite{ell2}. The strong inter-cluster (or inter-layer, polyhedra or ring), VW interactions in the medium range  are mediated by non-bonded lone-pair electrons.

The presence of short and medium range order (MRO)  in the solid gives rise to an internal structure made of inter-connected molecular clusters. Based on the range of dominant interactions, there are two natural length scales which dictate the behavior of the system. The first length scale comes from the short range molecular interactions within a block while the second originates from the collective degrees of freedom, of both phonon as well as non-phonon types. The latter can be regarded as the response of a molecular cluster, referred here as a block,  as a single unit to environment e.g the presence of other similar clusters.   The lack of long range order in amorphous systems clearly suggests that the block-block interactions should not be ''too'' long range (and can't be short either). Indeed the interaction of blocks with phonon strain field  generates a mutual RKKY type interaction (or virtual phonons) which is of inverse cube type \cite{vl, lg1}. The inter-molecular forces (i.e VW forces between two identical blocks with their centers at a distance $R$) at these length scales are also of long range type with $1/R^n$ spatial dependence, $n \le 3$ \cite{ajs,bb1}.  The dominance of intra-block interactions over (competition with) block-block (or inter-block) interactions  gives rise to  a coupled generic block structure  in the solid and is the main basis of our approach. As discussed later, an analysis of the interactions can then give us an estimation of the size of a basic block and is one of the primary objectives in this paper.

Our approach, discussed here as well as in \cite{bb1, qc1}, is  
based on a description of macroscopic glass solid as a set of interacting generic blocks subjected to phonon mediated inverse cube coupling. A similar approach was discussed in \cite{vl, dl}  but there are important differences  (i) the block in  our analysis are naturally arising molecular clusters;  contrary to \cite{dl}, the size of our blocks is fixed and is of medium range order in glasses (MRO), (ii) our approach  takes into account the molecular interactions within a single block too. An important result indicated by our present analysis is a universality related to the ratio $R_0/R_v$ of two crucial  length scales related to molecular interactions (hereafter referred as ''length-ratio''); here $2 R_0$ is the length scale  at which unretarded dispersion forces between two molecules are balanced by the phonon mediated coupling of their stress fields and $2 R_v$ is the distance between two nearest neighbor molecules interacting by VW interactions. 
As discussed in \cite{qc1}, the ''length-ratio'' also appears in the mathematical formulation of the average ultrasonic attenuation coefficient $Q^{-1}$ for elastic waves; the universality of the former for a wide range of materials than indicates the same for the latter and vice versa. In \cite{qc1}, a comparison of theoretical results  based on our formulation with experimental data for 18 glasses confirmed the universality of $Q^{-1}$. The present study confirms the same for the ''length-ratio''.

As discussed later in the paper, the ''length-ratio'' also leads to universality of the ''strength ratio'' ${\gamma_l \over \gamma_t}$ with $\gamma_{l,t}$ as the strength of the phonon mediated interactions between two molecules. 
In past, assuming the TTLS model as the correct theoretical description with $\gamma_{l,t}$ as the TLS-phonon coupling constants, the Meissner-Berret experiment analyzed the resonance energy absorption  from the acoustic field for many systems \cite{mb}.  Treating $\gamma_l, \gamma_t$ as the adjustable parameters, the experiment \cite{mb}  revealed that, below $T< 1^o K$, the ratio $\gamma_l \over \gamma_t$ is almost a constant between 1.5. and 1.6 for a wide variety of amorphous materials, irrespective of their chemical composition and molecular structure. The universal value of the strength-ratio, referred also as the Meissner-Berret (MB) ratio, contradicts the arbitrariness assumed in TTLS model and requires new reasoning \cite{mb}. This motivated a recent theoretical study  \cite{dl} to pursue the alternative model, based on a description of the macroscopic amorphous solid as a set of coupled generic blocks of arbitrary size in presence of external phonon strain field.  As the blocks are generic in nature, they can be regarded as the multi-level generalization of a TLS; (although contrary to blocks, TLS appear only as defects). The coupling parameters of a single block with transverse and longitudinal acoustic field are then expected to satisfy the universality of the MB ratio. This is indeed confirmed by the study  \cite{dl}, which theoretically derives, based on the concept of stress-stress susceptibility and a renormalisation approach, the resonance energy absorption by a single block as well as  macroscopic solid and expresses the MB ratio  in terms of another ratio, namely the ratio ${c_l \over c_t}$ of longitudinal and transverse sound speeds  in the solid. The available experimental data for the latter can then be used to calculate the MB ratio which indeed turns out to be ~1.5-1.6. The approach in \cite{dl} however is based on phonon mediated interactions of generic blocks only,  does not take into account the role of interactions within a block and as a consequence, does not explain the physical origin of MB universality.  This motivates us to pursue the present work which indicates the origin to lie in a competition between dispersion interactions and phonon mediated coupling between molecules. 

The success of our approach  in providing  theoretical justification of  the universality of the specific heat of MRO scale samples \cite{bb1} and ultrasonic attenuation for large samples encourages us to analyze its applicability to peculiarities of vibrational spectrum too.  Here we present theoretical conjectures regarding connection of boson peak frequency and Ioffe-Regal frequency  to MRO scales and compare the prediction  with available experimental data. This however requires a rigorous analysis and we expect to report detailed technical support for our conjectures in near future.

The paper is  organized as follows. Our primary focus in this work is to  present theoretical as well as numerical justification in support of the constancy of the length ratio ${R_0\over R_v}$ and thereby the BM ratio ${\gamma_l \over \gamma_t}$. Section II  compares the  interactions between molecules at short length scales with those emerging due to collective molecular  dynamics at large length scales and derives the  formulation for $R_0$ in terms of the molecular parameters. As discussed in section III, $R_0$ also turns out to be the size of the  basic block. The relation of the length-ratio with BM-ratio is discussed in section IV. The theoretical conjecture regarding relation of $R_0$ with characteristic vibrational frequencies is presented in section V.  A comparison of all the results obtained are also displayed in tables and figures. We conclude in section VI with a brief discussion our main results.

\section{Role of interactions: relevant length scales}

The radius, say $R_0$ of the dispersion forces in glasses beyond which $r^{-3}$ interaction becomes dominant can then be obtained by  a comparison of these interactions by two routes, those between  two molecules or between two clusters. 
As discussed below, both these approaches are needed   to formulate $R_0$ in terms of known molecular properties.

\vspace{0.2in}

\subsection{ Interactions between two molecules}

\vspace{0.2in}

Consider two molecules with their centers at a distance $r$ in a glass solid.  The existence of long wavelength  phonons at low temperatures leads to a phonon-mediated pair-wise interaction, decaying as inverse cube of distance between them. The corresponding  interaction energy is  
\begin{eqnarray}
V_{stress}(r) \approx  {\gamma_m^2 \over \rho_m \; c^2 \; r^3}
\label{vs1}
\end{eqnarray}
with $\rho_m$ as the mass-density of the basic block, $c$ as speed of the sound waves in the block  and $\gamma_m$ as  the strength of the phonon induced $r^{-3}$ coupling of  the two molecules. Note, as discussed in {\it appendix C}, $\gamma_m$ can be expressed in terms of the number of molecules and basic block parameters.

But as discussed in detail in \cite{bb1}, two molecules at a distance $r$ are  also acted by $r^{-6}$ type  interaction due to dispersion forces. Assuming the molecules to be in their ground state (valid at low temperatures), the interaction energy, averaged over all possible orientations, can be given as \cite{isra}
 \begin{eqnarray}
V_{dispersion}(r) \approx { C_6 \over   r^6} 
\label{vs2}
\end{eqnarray}  
with $C_6$ as the strength of  dispersion interaction between two molecules.

Our interest is now to find the distance $r=2 R_0$ at which the magnitude of the two interactions  is  equal.   This implies 
\begin{eqnarray}
V_{stress}(2 R_0)  &=&  V_{dispersion}(2 R_0) 
\label{vss}
\end{eqnarray}

Substitution of eq.(\ref{vs1}) and eq.(\ref{vs2}) in eq.(\ref{vss}) gives 

\begin{eqnarray}
R_0^3 = {\rho_m  \; c^2 \; C_6\over   8 \gamma_m^2}.
\label{r03}
\end{eqnarray}

As the Hamaker constant $A_H$ is a constant for materials, it is better to express $C_6$ in terms of $A_H$: $C_6 \approx {A_H \over \pi^2 \;  \rho_n^2}$, with $\rho_n={1\over \Omega_{\rm eff}}$ as the number density of the molecules and  $\Omega_{\rm eff}$ as the average volume available to a typical molecule. (Note here the number of those molecules should be considered which interact through dispersion interaction). 
Consider a typical molecule of radius $R_m$; its molar volume can then be expressed as 

 \begin{eqnarray}
\Omega_{m} = s_m \; R_m^3 = {M \over \rho_m N_a}
\label{ome0}
\end{eqnarray}
with $M$ as the molar mass of the particle interacting by VWD, $N_a$ as the Avogadro number and 
with $s_m$ as a structure constant  e.g. $s_m =4 \pi/3$ assuming a spherical shape for the molecule.
The effective volume available to each molecule in the system however also depends on their packing. Assuming  $2 R_v$ as the average distance between two neighboring molecules  in the system, the average volume, say $\Omega_{\rm eff}$, available to a typical molecule is 

\begin{eqnarray}
\Omega_{\rm eff} = s_m \; (R_v +R_m)^3 \approx  (1+y)^3 \; \Omega_m
\label{ome}
\end{eqnarray}
where $\Omega_m$  is the molar volume: $\Omega_m={M \over \rho_m N_a}$  The last equality in the above equation follows by taking $\Omega_m = s_m \;  R_m^3$. Further writing  $R_v = y \; R_m$ along with eq.(\ref{ome}) in the definition of $C_6$ gives

\begin{eqnarray}
C_6 \approx {A_H \; (\Omega_{\rm eff})^2 \over \pi^2} \approx {{ s_m^2 \over \pi^2} \;(1+y)^6 \; R_m^{6} \; A_H}
\label{c6}
\end{eqnarray}

The above information can now be used to rewrite $R_0$ in terms of known molecular parameters. 
Substitution of eqs.(\ref{ome0}-\ref{c6}) in eq.(\ref{r03}) leads to

\begin{eqnarray}
R_0^3 = {(1+y)^6 \;  c^2 \; A_H  \; M \; \Omega_m \over    8 \;  \pi^2  \; \gamma_m^2 \; N_{av}}.
= {(1+y)^6 \;   \; A_H  \; M^2 \; c^2 \over    8 \;  \pi^2  \; N_{av}^2  \; \gamma_m^2  \; \rho_m}.
\label{r05}
\end{eqnarray}

Clearly $R_0$ is an important length scale, related to equivalence  of the above two interactions energies, one  short and other  long range; for $r < R_0$, dispersion force dominates and for  $r > R_0$, phonon mediated coupling dominates. As discussed below, the competition leads to  organisation of molecules in structural sub-units i.e basic blocks of linear size $R_0$.

\vspace{0.2in}

\subsection{ Interactions between two neighboring  clusters}

\vspace{0.2in}

Consider a cluster of molecules within a sphere of radius $R_0$, the dominant energy of interaction between any two molecule within the sphere  is that of dispersion. But for two surface molecules diagonally opposite (i.e distance $2 R_0$ across a diameter), the two  interaction energies  are equal. For any molecule outside the spherical surface, the phonon mediated coupling dominates over  that of  dispersion.  This clearly indicates existence of structures of linear size $2 R_0$.  But the question is, if one considers the collective interactions of a cluster of molecules  with another neighboring cluster, for what size of the clusters, the two interaction energies still be equal?

 Consider two spherical blocks of radius $t$ with $2 R_v$ as the closest separation distance between them, each containing $g_0$ molecules and subject to following conditions,

(i) each block is dispersion dominated i.e within one block, the dominant force between any two molecules is that of dispersion.   

(ii) The size of the blocks are so chosen that, for neighboring blocks, dispersion interaction   between them is equal to that of phonon mediated coupling between them.
 
 The phonon mediated coupling between the clusters, with their centers separated by a distance $2(t+R_v)$,  can be expressed as (for $t > R_v$) (derived in {\it appendix} A)

\begin{eqnarray}
V_{b, stress}(t)  
&\approx & - {4 \; \pi^2  \; \gamma_m^2 \; \rho_n^2  \over  3 \; \rho_m \; c^2 } \; t^3
\label{vb2}
\end{eqnarray} 
with $\rho_n$ as the number density of the interacting particles.

Here subscript $b$ refers to interaction energy of two neighboring basic blocks. 
 Further  the dispersion interaction between two spherical clusters of radius $t$  separated by a contact distance $2 R_v$ can be given as \cite{ham}

\begin{eqnarray}
V_{b, dispersion}(t)  = - {\pi^2 \; \rho_n^2 \; C_6 \over 24 } \; { t \over  R_v}
\label{vb1}
\end{eqnarray}

Let us now impose the condition (i) that
\begin{eqnarray}
 V_{b, dispersion}(t) = V_{b, stress}(t)
 \label{v1}
 \end{eqnarray}

Substitution of eq.(\ref{vb1}) and eq.(\ref{vb2}) in eq.(\ref{v1}) then leads to 

\begin{eqnarray}
{t \over R_v} \approx  {32  \; \gamma_m^2 \; t^3  \over  \rho_m \; c^2 \; C_6} \; 
\label{vbn3}
\end{eqnarray}

Now using eq.(\ref{r03}) to replace $C_6$ in the above,  the size $t$ of the basic block can be expressed in terms of $R_0$ and $R_v$,

\begin{eqnarray}
 t^2 =  { R_0^3\over 4 \; R_v}
 \label{ve0}
 \end{eqnarray}

Note as $R_0 \ge 4 R_v$, (the total distance between the centers of two neighboring molecules being $2(R_m+ R_v)$, each assumed spherical with radius $R_m$,   and $R_v= y \;  R_m$ with $y$ a material dependent constant of order 1), this along with  the above implies

\begin{eqnarray}
 \left({t\over R_0}\right)^2   = { R_0\over 4 \; R_v}  \qquad  \ge 1 
 \label{ve1}
 \end{eqnarray}

It is worth emphasizing here that, for two neighboring blocks each of radius $t > R_0$, the two interaction strengths can not be equal and $V_{b,stress}$ will always dominate $V_{b,dispersion}$. This can directly be seen from 
eq.(\ref{vb1}) and eq.(\ref{vb2}) which leads to ${V_{b, stress}(t)\over V_{b, dispersion}(t)}={4 R_v t^2 \over R_0^3}$ for arbitrary $t$. Clearly with $R_0 > 4 R_v$ and $ t > R_0$, ${V_{b, stress}(t) > V_{b, dispersion}(t)}$.

To get an idea of as to which interactions are really influencing the dynamics at different length scales, it is  instructive to analyze their order of magnitude: typically $C_6 \approx 434 \; A_H R_m^6 \sim 10^{-18} \; R_m^6$ ( from eq.(\ref{c6}) and $A_H \sim 10^{-20} \; J$ from Table 1) and ${\gamma_m^2 \over \rho_m c^2} \approx 10^{-49} \; J. m^3$ (from Table 1). 
For two molecules at a distance $2 R_0 = 8\; R_m$ with $R_m \sim 3 \; \AA$, eq.(\ref{vs1}) and eq.(\ref{vs2}) then give $V_{stress}(2R_0) = V_{dispersion}(2R_0) \approx 10^{-23} \; J$ which is of the order of $k_B T$, the thermal energy  needed  at  temperature $T \sim 1^o {\rm K}$ (with $k_B$ as the Boltzmann constant). The interaction energy  for two neighboring blocks with their centers separated by a distance $2(R_0+R_v)$ is also of the same order:   eq.(\ref{vb1}) and eq.(\ref{vb2}) give $V_{b, stress} = V_{b, dispersion} \approx 10^{-23} \; J$.  This is however not the case for two molecules at minimum possible distance $2(R_m+R_v) \approx 4 R_m$, then  $V_{dispersion} \sim 10^{-17} \; J$ and $V_{stress} \sim 10^{-22} \; J$.

\vspace{0.2in}

\section{Size of a basic block}

The behavior of a macroscopic glass solid can be analyzed in terms of a collection of molecular clusters i.e blocks   of arbitrary shape.
As the universal properties are expected to originate from the length scales beyond short range order (SRO), the linear size of each basic block  can be chosen to be large enough compared to the typical inter-atomic distance  but is otherwise arbitrary. At this stage, we have two options for the upper size limit: {\it first option} is that the size is comparable to the range of unretarded dispersion forces among molecules which dominate within a block. Due to orientation free nature of these forces, the interaction among the molecules within a basic block can be regarded as isotropic. The stress fields of two different  blocks are however coupled mediated by phonons \cite{qc1,vl,dl}. The {\it second option} is that we leave the upper limit arbitrary; the molecules within a block in that case  interact via phonon mediated coupling (dependent on inverse cube of the distance between molecules) too.  
One could refer the basic blocks of {\it first} and {\it second} options as  {\it dispersion} blocks and {\it stress} blocks.

In principle, proceeding with either option should lead to analogous results for the physical properties; (the 2nd option was adopted in \cite{vl,dl} for a renormalisation approach to analyze universality). In practice however it is easier to choose the first one for following reasons 
(i) the dispersion forces being the short range one, they correspond to more natural choice to define the building block size.
(ii) isotropic nature and homogeneity of the interaction within the block is a better approximation in this case which is technically helpful to calculate the level density and other properties. 
(iii) as discussed later, it gives us the route to calculate the coefficient of ultrasonic attenuation of phonons  in terms of molecular properties and confirm its universality at nano-scales.

\vspace{0.2in}

Let us consider the whole glass solid divided into spherical blocks of radius $t$ with $2 R_v$ as the closest separation distance between them.  Here the choice of a spherical shape is natural, keeping in view the medium range order  in glasses as well as molecule shape assumed to be spherical; it also helps in technical simplification without any loss of generality.  Proceeding with the first option, it is clear that the block size should not be big enough to allow the onset of  phonon-mediated $r^{-3}$ interactions. 

As indicated by eq.(\ref{ve1}), the balancing of two types of interactions between neighboring blocks leads to an inequality relation between $t$ and $R_0$. 
Now consider  two molecules  within a single spherical cluster of radius $t$; the maximum possible distance between the molecules is then $2 \; t$. The appearance of a block type structure now requires  that the $r^{-6}$ interaction between them  is more dominant than that of the $r^{-3}$ interaction.   This implies 
\begin{eqnarray}
V_{stress}(2 t) \quad  \le \quad  V_{dispersion}(2 t)  \label{vs3}\end{eqnarray}
Using eqs.(\ref{vs2}, \ref{vss}), we have 
\begin{eqnarray}
t^3   \quad \le \quad   { 8 \; C_6 \;  \rho_m \; c^2 \over 64 \gamma_m^2} 
\label{vsz}
\end{eqnarray}
This along with eq.(\ref{r03}) now gives the condition
\begin{eqnarray}
\left({t \over R_0}\right)^3  \le  1    
\label{vsv3}
\end{eqnarray}
To satisfy both eq.(\ref{ve1}) and eq.(\ref{vs3}) simultaneously, one must have ${t \over R_0 }  =  1$ and thereby
\begin{eqnarray}
 t  = R_0 \approx 4 \; R_v
\label{vb4}
\end{eqnarray}


%

The above information along with eqs.(\ref{ome0}, \ref{ome}, \ref{r05}) further gives the number  $g_0$ of the molecules within each block as

\begin{eqnarray}
g_0 =  {\Omega_b \over \Omega_{\rm eff}} \approx {1 \over (1+y)^3} \; \left({t \over R_m}\right)^3 =  {y^3 \over 8 \; (1+y)^3} \; \left({R_0 \over R_v}\right)^{9/2}   
\label{g0}
\end{eqnarray}

\vspace{0.1in}

 The result (\ref{vb4}) is one of the central results of this paper and can intuitively be explained as follows: as $R_0$ defines the length scale at which the relevant interaction energies between two molecules become equal,  it should clearly depend on the smallest distance $R_v$ accessible between the molecules which are interacting by the dispersion forces. As $R_v$ varies from one material to another, $R_0$ is also expected to vary. Further following short range of dispersion forces, $R_0$ is expected to have  a linear dependence on $R_v$.

 As eq.(\ref{vb4}) indicates, the ratio  ${R_v \over R_0}$ is independent of the glass material and, as discussed later in sections IV and V, it leads to a material independent values of the MB Ratio ${\gamma_m \over c}$ as well as the ultrasonic attenuation coefficient $Q^{-1}$ (see \cite{qc1} for the latter).

The results in the next section are based on the relation (\ref{vb4}); before proceeding further, it is therefore desirable to verify it numerically. 
For numerical calculation of $R_v$, we use  $R_v = y R_m$ with $R_m$ given by eq.(\ref{ome0}). As eq.(\ref{r05}) is obtained without using the relation (\ref{vb4}), it can therefore be used to calculate $R_0$.  This requires a prior knowledge of  $A_H$, $y$ and the molar mass $M$. The values of $A_H$ are taken from \cite{qc1}). As the quantitative information about $R_v$  available for a wide range of materials suggests $R_v \sim R_m$, we use $y=1$ for our calculations. (In general $y$ fluctuates from one glass to another with $y \approx 1$ as its average value. The results given in table 2 are based on $y \approx 1$ but a better quantitative result can be obtained by taking glass-specific values for y). With $M$ as the mass  of the molecular unit undergoing VWD interaction, we consider two alternatives:  (i) mass of the basic structural unit which dominates the structure of the glass and participates in the dispersion interaction (later referred as vwd unit), or, (ii) the mass of the formula unit of the glass (later referred as formula unit); (here, for example for $SiO_2$ glass, $SiO_2$ is the formula unit but dominant structural unit can be $SiO_4$ or $Si(SiO4)$, see {\it appendix A} of \cite{qc1} for details).  Clearly, with dispersion interaction as the basis of our analysis,  $1^{st}$ option seems more appropriate. This is also confirmed by 
our numerical analysis indicating the theoretical predictions based on $M_1$ closer to experimental results.

The numerical result for $R_m$, $R_0$ and $t$ for 18 glasses, obtained by eq.(\ref{ome0}), eq.(\ref{r05}) and eq.(\ref{ve0}) are displayed in tables II and III (with parameters required for their calculation in table I); here the labels  $R_{0l}, t_l$ and $R_{0t}, t_t$ refer to the values corresponding to longitudinal and transverse sound velocities. As theoretically predicted, the $t_l$ values are close to $R_{0l}$ and $t_t$ to $R_{0t}$. Furthermore $2 t_l$ and $2 t_t$ values for all 18 glasses are of typical MRO length scales. Note  ideas indicating relevance of MRO length scales were suggested in past too e.g. \cite{du, vdos, ell3, mg}, and, as displayed in table II, our $R_0$ is approximately the same as $R$ of \cite{ell3} (see table I of \cite{ell3}).  

The results are also  illustrated in figures 1 and 2 which indicate that the relation in eq.(\ref{vb4}) is better supported by the mass  $M_1$. This is also consistent with the results for other low temperature properties  of glasses \cite{qc1,bb1}.  As figure 1 indicates, ${R_0 \over R_m} \sim 4$ and $R_0 \approx t$; comparison of these results with our theoretical predictions $R_0 = 4 R_v = t$ is also consistent with our assumption $R_v \approx R_m$.

\section{Ratio of molecular coupling strengths}

With relation (\ref{vb4}) numerically validated, it can now be used to derive 
the ratio ${\gamma_m \over c}$. The steps are as follows.
Substituting eq.(\ref{vb4}) in the left side of eq.(\ref{r05}), along with $R_v = y \;  R_m$, gives 
\begin{eqnarray}
R_m^3 = {\rho_m  \; c^2 \; C_6\over   512 \; \gamma_m^2 \; y^3}= {512 \; \pi^2 \over s_m^2 \; \rho_m  \; A_H}  \; { y^3 \over (1+y)^6} \; \left( {\gamma_m \over c} \right)^2
\label{r06}
\end{eqnarray}

 A comparison of eq.(\ref{r05}) with $R_m^3 ={M \over s_m \; \rho_m \; N_{av}}$  (see  eq.(\ref{ome0})) then leads to 
\begin{eqnarray}
 {\gamma_m \over c}  = \left({s_m \; M  \; A_H \over 8 \;\pi^2 \;  N_{av} }  \right)^{1/2} \; {(1+y)^3\over 8 \; y^{3/2}}
\label{r006}
\end{eqnarray}
with  right side in the above equation dependent on the molecular parameters, it remain same for both longitudinal as well as transverse propagation of sound waves. Using $\gamma_m=\gamma_l, \gamma_t$ and  $c=c_l, c_t$, this leads to

\begin{eqnarray}
{\gamma_{l} \over c_l} ={\gamma_{t} \over c_t} \qquad \rightarrow \qquad {\gamma_{l} \over \gamma_{t} } = {c_l \over c_t}
 \label{bm1}
\end{eqnarray}
The velocity ratio ${c_l \over c_t}$ is experimentally observed to vary very little for a wide range of amorphous material (${c_l \over c_t} \sim 1.5-2.0$) \cite{mb,dl}. Eq.(\ref{bm1})  then implies a similar behavior for the ratio ${\gamma_{l} \over \gamma_{t} }$ too.

Eq.(\ref{r006}) along with $M$ (for both choices $M_1, M_2$ (see \cite{qc1} for detail) and $A_H$ (given in table I)  leads to theoretically predictions for the ratio ${\gamma_m \over c}$. The results for 18 glasses based on eq.(\ref{r006}) are displayed in  table IV, for both $M_1$ and $M_2$, along with available experimental data for ${\gamma_l \over c_l}, {\gamma_t \over c_t}$ (taken from \cite{mb} and \cite{plt}). A comparison of theory and experimental results, illustrated in figure $3$ for $M_1$ and figure 4 for $M_2$, respectively indicates the validity of eq.(\ref{r006}). As clear from the table, here again the results for $M=M_1$ are closer to experimental data, thus indicating the molecules interacting by VWD interaction as an appropriate choice for the present analysis. This  is also consistent with our theoretical approach assuming  VWD interactions as the relevant interaction for length scales less than MRO. 

It is worth comparing the above result with previous theories. Based on TTLS model, the existence of relation in eq.(\ref{bm1}) was first considered in  \cite{mb}; here $\gamma_l$  and $\gamma_t$ were obtained as adjustable parameters to experimental acoustic data (see {\it appendix A} for a brief review). 
Later on, based on generic block model, a detailed theoretical investigation of the coupling strengths was carried out in \cite{dl} but as mentioned in section I, the study did not explicitly consider the role of interactions within a block.  Contrary to above  models,  an additional interaction persists  within a block in our model i.e VW forces  dominant at short length scales among molecules. (In TTLS model, the TLS atoms being defects, their distribution is not  dense enough to take into account the short range inter-atomic forces). The important aspect of our approach is that it not only confirms  the constant value of ratio ${\gamma_l \over \gamma_t}$ among amorphous materials but also reveals how a conspiracy between two material properties results in material-independent value of ${\gamma_a \over c_a}$. Note as displayed in table III, ${\gamma_a \over c_a}$ is nearly a constant for $18$ glasses. The latter is turn indicates a relation between the molar mass $M$ and Hamaker constant $A_H$ both material dependent.

\section{Connection to characteristic  frequencies}

With length scale ${R_0}$ appearing as the back bone of many glass properties within our approach e.g. specific heat \cite{bb1}, internal friction \cite{qc1}, strength ratio etc,  it is natural to query whether there exist any relation of $R_0$ with peculiar features of the low frequency vibrational spectrum.

Based on experimental observations in past many decades,  the vibrational spectrum in amorphous systems, in the energy range $2 \to 10 \; meV$, deviates significantly from phonons  and shows an excess density of states (DOS) relative to the one based on Debye's theory ($\rho(\omega) \sim \omega^2$). Also confirmed by recent experiments \cite{lern, mizu}, the vibrational density of states (VDOS) $\rho(\omega)$  changes with increase in frequency $\omega$, from $\omega^4$ (for $\omega < \omega_{bp}$) to $\omega^2$ (for $\omega > \omega_{bp})$ with $\omega_{bp}$ as a frequency characteristic of material. Further the form of reduced density $\rho(\omega)/\omega^2$ displays a peak at $\omega=\omega_{bp}$, referred as Boson peak (BP). For $\omega > \omega_{ir}$, however, the vibrational transport is  diffusive \cite{allen}; here $\omega_{ir}$ is another characteristic frequency of the material at which Ioffe-Regel (IR) criterion $l \sim \lambda$ is satisfied by phonons, with $l$ as the  mean free path of phonon and $\lambda$ as its wavelength. The strong scattering in this limit leads to localization of phonons, rendering them ill-defined and incapable of heat-transportation.

The experimental evidence indicating diffusive vibrational transport above the IR limit $\omega > \omega_{ir}$ has motivated consideration  of diffusions  (contrary  to ballistic transport by phonons) \cite{bkp,degi,allen}. The latter were identified by diffusivity calculations in some real glasses using molecular dynamics approach  and also in granular jammed systems with repulsive forces between the particles.  A simple model analysing transport in terms of diffusions was suggested in \cite{bkp} with $\omega_{ir}$ defined as the IR frequency at which crossover between phonons and diffusions occurs. Such a crossover was confirmed by many experiments and $\omega_{ir}$ was found close to boson peak frequency $\omega_{bp}$. This again led to many debates, inconclusive so far, regarding the relation of $\omega_{bp}$ with $\omega_{ir}$. 

A physical explanation for diffusions within our approach can intuitively  be suggested as follows. For $T > 20^o K$,  the orientational disorder of the induced dipoles at MRO scale or less affects the phonon dynamics causing their scattering and thereby localization.
As mentioned in section I and III (and in detail in \cite{qc1}), the block subunits within a sample of macroscopic size interact with each other via phonon mediated interactions of their stress fields. The localization of phonons above $T > 20^o K$ however may weaken the coupling between basic blocks and it may not even be well defined  for faraway blocks. Nonetheless the neighboring blocks can still interact with each other by local density fluctuations, thus leading to diffusion as in the case of electronic transport in a three dimensional disordered lattice. This is however just an speculation at this stage and need to be confirmed by technical analysis e.g of inverse participation ratio of the many body product states of the basic blocks.

Based on  the above idea, the relation of $\omega_{ir}$ to $R_0$ emerges as follows. 
 As the minimum value of phonon mean path is determined by the size of the entity (i.e a basic block in our case) causing the scattering, this implies $l \sim t \approx 2 R_0$. However experiments indicate $l \sim \lambda$ for $T > 20^o K$.  The above then gives 
\begin{eqnarray}
\omega_{ir,a} \sim {c_a \over 2 t} \approx {c_a \over 2 R_0}
\label{omd}
\end{eqnarray} 
with $c_a=c_l,c_t$ as the longitudinal or transverse sound velocity in the medium.
The table V lists the predictions for $\omega_{ir,a}$ based on the above prediction.
   As displayed in the table, the values $\omega_{ir,t}$ (obtained from either $R_{0t}$ or $t_t$) are closer to available experimental data for $\omega_{bp}$. This is consistent with past studies indicating dominant role of transverse phonons in heat transport (e.g. \cite{wang}) and also provides evidence supporting $\omega_{ir,t} \sim \omega_{bp}$ in amorphous system \cite{shin, bfpt} (also see \cite{nie} in this context).

The form of Boson peak  is known to be universal for  glasses irrespective of their varying chemical composition and short range order; its physical origin therefore has to lie in features common in all glasses and based on general physical principles. 
Current understanding of the origin of BP can  be classified in two broad categories: (i) the excitations  giving rise to the BP are non-phonon type, believed to arise due to structural disorder e.g. fractons \cite{frac,gg1}, quasi-localized vibrations \cite{spm,paras},  Euclidean random matrices \cite{grig} or structural correlations over $10 \to 20 \; \AA$ \cite{du, vdos,ell3,sksq,mg}, and, (ii) the short range order being same in both glass and crystal, VDOS of the amorphous system is just a modification of that of a crystal (e.g. due to a random fluctuation of force constants \cite{sdg,schi}), with the BP as the broadened version of Van Hove singularity (a typical feature for excitations in a periodic structure) \cite{tara,chum}. The question whether BP originates from  the structural disorder  or the short-range order of the glass (the glassy counterpart of the first (transverse) Van Hove singularity (VHS) in crystal) has led to contradictory claims e.g. \cite{chum, wang,nie} and still remains controversial.

Although the present work is not an attempt to settle this controversy,
the vibrational excitations giving rise to BP in our approach clearly do not arise from short range order (i.e of the range of chemical bonds) but rather from medium range order in  amorphous systems. However the structural disorder in our case appears only in the form of randomly oriented instantaneous dipoles interacting by VW forces. 
Following physical scenario seems to emerge intuitively from our analysis: the amorphous  sample of experimental size is basically a collection of MRO sized molecular clusters i.e basic blocks, forming regular partitions of the former. (The sample can also be viewed as a regular  ''lattice'' with each site occupied by a basic block). At  temperature $T < 20^o K$, the blocks seemingly  respond to transport as an array of periodic structures, subjected to phonon mediated coupling.  This leads to  resonant scattering of phonons by the blocks which in turn ensures large mean free paths, thereby reducing the attenuation; the latter is confirmed by a detailed calculation in \cite{qc1}. Further, as explained in section V of \cite{bb1}, the DOS of an amorphous  sample of experimental size is a convolution of the DOS of the basic blocks. The form of the latter changes at an energy scale (discussed in section III of \cite{bb1})
\begin{eqnarray}
e_0 = {\sqrt{2}\over 3 b \gamma} \approx  {4\sqrt{6} \; A_H\over 27 \gamma} \left({1+y\over y}\right)^{9/2}  \approx 10^{-21} \; J
\label{e0}
\end{eqnarray}
with $y={R_v\over R_m} \sim1$, $\gamma=\sqrt{2} .3^{2 g_0/3}$ and $b$ defined in section II.E of \cite{bb1}; this intuitively suggests  $e_0$ in the same energy range as  the boson peak
\begin{eqnarray}
\omega_{bp} \sim {e_0 \over \hbar} 
\label{omb}
\end{eqnarray} 

To check the above conjecture, we analyze  $e_0$ values for 18 glasses with $b$ values taken from \cite{bb1}.  The table V lists the $e_0$ values based on eq.(\ref{e0}) as well as known experimental values for $\omega_{bp}$; the two values  differ by a factor of $2$  $3$. This is expected because while the $e_0$-values correspond to edge-bulk meeting point of the VDOS of nano-size samples, the experimental values of $\omega_{bp}$ refer to macroscopic ones. A resolution of this issue requires an analysis of the convolution of the vodos of basic blocks; we expect to report it in near future.

\section{Conclusion}

In the end, we summarize our main ideas and results.

A competition of unretarded dispersion interaction between two molecules with  phonon mediated coupling of their stress fields gives rise to a natural length scale $R_0$  which is theoretically predicted to be a constant independent of the glass material (in units of $R_v$, the distance of  minimum approach between two neighboring molecules); the prediction is verified numerically for $18$ glasses. This also reveals a dominance of different interactions at short and large scales:  the dispersion interaction within molecular clusters of linear size $R_0$ and a inverse cube interaction at larger scales.  This competition in turn leads to  an organisation of the molecules in structural sub-units i.e basic blocks of linear size $R_0$. The standard formulations of the interaction energies,  then give $R_0$ in terms of the molecular parameters. Based on $18$ glasses, we find $R_0$ to be of the order of well-known MRO length scales in glasses.

Based on constant value of length ratio ${R_0 \over R_v}$, our approach also  provides a theoretical pathway to calculate the coupling strength ratio ${\gamma_l \over \gamma_t}$ of the longitudinal and transverse coupling strengths of molecules at ultra low temperature.  Our analysis also reveals that ${\gamma_l \over \gamma_t}$ is  same as (i) the velocity ratio ${c_l \over c_t}$, and, (ii) the experimentally observed universal value for the  corresponding ratio in case of TLS i.e MB ratio \cite{mb}. A comparison with experimental data for 18 glasses supports our prediction. The omnipresence of dispersion forces indicates the application of our results to other disordered materials too.

The physical insight based on our theoretical analysis in \cite{bb1}  encourages us to predict the  relation of $R_0$ to boson peak frequency and Ioffe-Regel frequency of the vibrational spectrum at low temperature in amorphous systems. As our approach is based on VW forces, existing in all materials, its generalization to  may explain the peculiar features e.g. boson peak common in both non-crystalline as well as disordered crystalline materials.

An important point worth indicating in the end is following. The numerical results obtained here are based on the approximation $R_v \sim R_m$. In general, the ratio $R_v/R_m$ is expected to fluctuate from one glass to another. This may explain the fluctuations of experimentally observed values around an average value,  instead of being a constant as theoretically predicted by us for amorphous systems.

\newpage 

\appendix

\section{Derivation of eq.(\ref{vb2})}

The derivation follows the similar steps given in section 1 and 2 of  \cite{ham} for the derivation of eq.(\ref{vb1}).

The energy of interaction, say $V_{b,stress}$, between two spheres , both say of radius $R_0$ and  volume $\Omega$ and each  containing $\rho_m$ molecules per unit volume is given by

\begin{eqnarray}
V_{b, stress}  
 = - \int_{\Omega} {\rm d} v_1 \int_{\Omega} {\rm d} v_2  \; {\rho_m^2 \; \lambda \over r^3}
\label{vba1}
\end{eqnarray} 
with ${\rm d}v_1, {\rm d}v_2$  as the volumes elements at a distance $r$.  Note here $\lambda= {\gamma^2 \over \rho_m c^2}$. 

The only difference between our case and Hamaker's is that instead of $1/r^6$ interaction, here we have $1/r^3$. To simplify calculation, here we take the same radius of both speheres (i.e $R_1=R_2=R_0$)
Following exactly the same steps as in \cite{ham} to derive eq.(6) for $E$, here we have (with $C= 2 (R_0 + R_v)$)

\begin{eqnarray}
V_{b,stress}= - {\pi^2 \rho_m^2 \lambda \over C} \; \int_{C-R_0}^{C+R_0} {\rm d} R  \; (R_0^2 -(C-R)^2) \;  \int_{R-R_0}^{R+R_0} {\rm d} r  \; { (R_0^2 -(R-r)^2)\over r^2}
\label{vba2}
\end{eqnarray} 

The 2nd inetgral gives 
 
 \begin{eqnarray}
  \int_{R-R_0}^{R+R_0} {\rm d} r  \; { (R_0^2 -(R-r)^2)\over r^2}
  &=& - 4 R_0 + 2 R \log {R + R _0\over R-R_0} \\
 &=& -4 R_0 + 2 R \left({2 R_0 \over R} +  {2 R_0^3 \over 3 R^3} + \ldots \right)  \\
 &\approx & {4 R_0^3 \over 3 R^2} + \ldots 
  \label{vba3}
\end{eqnarray} 

The substitution of eq.(\ref{vba3}) in eq.(\ref{vba2}) gives 
\begin{eqnarray}
V_{b,stress} &=& - {4 R_0^3 \over 3} \; {\pi^2 \rho_m^2 \lambda \over C} \; \int_{C-R_0}^{C+R_0} {\rm d} R  \; {(R_0^2 -(C-R)^2) \over R^2} \\
& = & - {4 R_0^3 \over 3} \; {\pi^2 \rho_m^2 \lambda \over C} \; 2 R_0
\label{vba4}
\end{eqnarray} 
Further using $C=2 (R_0 + R_v) \approx 2 R_0$ and substituting $\lambda= {\gamma_m^2 \over \rho_m c^2}$,  we have 
\begin{eqnarray}
V_{b,stress} =  -   {4 \pi^2 \rho_m \gamma_m^2 \over 3 \; c^2} \; R_0^3 
\label{vba5}
\end{eqnarray} 
which is same as eq.(\ref{vb2}).


\section{Coupling constant $\gamma_{l,t}$ in TTLS model}

With extensive research on the TTLS model in last few decades, it is discussed in many papers in great details. The generic block model without/ with consideration of dispersion forces is discussed in \cite{vl, dl} and \cite{bb1,qc1} respectively.  
To provide  a clear idea of our analysis and keep the paper self-content, here we revise some of the ideas and formulations for TTLS model. 

In TTLS model of the amorphous solid,  the atoms  occupying two adjacent minimas (referred as TLS or defects) in the amorphous solid are assumed to  undergo quantum tunnelling from one to other, leading to a splitting of the ground state of the molecule. 
The role of defects  in this model is played by the atoms  occupying two adjacent minimas (referred as TLS); the quantum tunnelling from one minima to other, leads to a splitting of the ground state of the molecule. 

The TTLS Hamiltonian consist of three terms: an elastic (phonon) term, a term corresponding to a set of non-elastic TLS and a term  corresponding to phonon-TLS coupling. 
%
The TLS are assumed to have a broad distribution of the energy splittings and relaxation times  and interact with propagating elastic waves by  the resonant as well as relaxational mechanisms. 
Assuming the linear coupling between a TLS and the phonon-strain field, the Hamiltonian $H$ for the interaction can be written as 

\begin{eqnarray}
H=H_0 + \gamma_a \; (k \; A \; {\rm e}^{i \omega t}) \; V.
\label{htls}
\end{eqnarray}
with $H_0={1\over 2}\begin{pmatrix}E & 0 \\ 0 & -E \end{pmatrix}$ and $V=
{1 \over 2 E} \begin{pmatrix}\Delta & \Delta_0 \\\Delta_0 & -\Delta\end{pmatrix}$.
Here $H$ is written in the eigenbasis of TLS with $E=\sqrt{\Delta_0^2+\Delta^2}$ as the energy-splitting between two states, $\Delta, \Delta_0$ as the diagonal and off-diagonal matrix elements of the TLS-phonon coupling with $\gamma_a$, $a=l,t$  as the coupling constants with the longitudinal and transverse phonon fields, respectively, of wavenumber $k$, amplitude $A$ and frequency $\omega$. The coupling parameters  $\Delta, \Delta_0$ and $\gamma_a$ can in general vary among TLS and only those in resonance with external phonon field $E=\hbar \omega$ can absorb energy which linearly increases with time $t$.  The resonant interaction of a TLS with phonons results in a relative velocity change $\left(\Delta c_a \over c_a \right) = {\mathcal C}_a {\ln}(T/T_0)$  and in an unsaturated absorption $\alpha_a = {\pi \omega {\mathcal C}_a \over c_a} \; \tanh\left({\hbar \omega \over 2 k_b T} \right)$ with $T_0$ as an arbitrary reference temperature and 
\begin{eqnarray}
{\mathcal C}_a= {{\overline P} \gamma_a^2 \over \rho_m c_a^2}
\label{ts}
\end{eqnarray}
 with $\rho_m$ as the mass density and ${\overline P}$ as the spectral density (number of TLS per unit volume and energy), $c_l, c_t$ as the longitudinal and transverse sound velocities. The relaxational interaction although contributes negligibly to velocity change but it affects absorption due to $1$-phonon process and asymptotically leads to $\alpha_a = {\pi^4 {\mathcal C}_a K_3 T^3 \over 96 v_a}$ with $K_3={4 k_B^3 \over \pi \rho_m \hbar^4} \left({\gamma_l^2\over c_l^5}+{\gamma_t^2 \over c_t^5}\right)$ \cite{mb}. 

Eq.(\ref{htls}) describes the coupling of a single TLS with the phonon field. The TLS are also subjected to phonon-induced ${1\over r^3}$ coupling. 
%
Treating ${\mathcal C}_a$ and $K_3$ as independent fitting parameters, experimental acoustic data can be well described by TTLS model. The fitted values are then used to determine the characteristic TLS parameters i.e ${\overline P}$ and $\gamma_a$ \cite{mb}.

\newpage

\begin{sidewaystable}[ht!]
\caption{\label{tab:table I} {\bf Physical parameters for 18 glasses:}
The table lists the available data for the physical parameters appearing in eq.(\ref{r05}) and eq.(\ref{r006}).
The $\rho, c_l, c_t, {\overline P}$ data from \cite{mb} (or \cite{plt} if not available in \cite{mb}) is displayed 
in columns $3^{rd}, 4^{th}, 5^{th}$ and $8^{th}$, respectively. 
 The columns $6^{th}$ and $7^{th}$ give the 
$\gamma_l$ and the $\gamma_t$ values, taken from  \cite{mb}  except for few cases; (for those marked by a star (*), the  values are obtained either from \cite{plt} or from $C_l, C_t$ values given in \cite{mb} along with eq.(\ref{ts}). The $A_H$ values  given in columns $9^{th}$ are taken from \cite{bb1, qc1}. The  molar mass values, referred as $M_1$  for the vwd unit along with its composition is given in columns $10^{th}$ and $11^{th}$ and the mass $M_2$ for formula unit (same as glass molecular weight) in column $12^{th}$ respectively. }
\begin{center}
\begin{ruledtabular}
\begin{tabular}{lccccccccccr}
Index & Glass & $\rho_m$  &  $c_l$ & $c_t$  & $\gamma_l$ &  $\gamma_t$   & ${\overline P}$  &  $A_H$  & $M_1$   & Vwd unit & $M_2$ \\
\hline
 &   & $\times 10^3 {Kg/m^3}$ &  ${km/sec}$ & ${km/sec}$ & $ev$ & $ev$ & $10^{45}/J. m^3$ & $\times  10^{-20} \; J$ & $gm/ mole$  & &  $gm/mole$ \\
\hline
  1 &  a-SiO2          &    2.20 &    5.80 &    3.80 &    1.04 &    0.65  &      0.8 &  6.31  &  120.09  & [$Si(SiO_4$] &      60.08 \\
    2 &  BK7             &    2.51 &    6.20 &    3.80 &       0.9 &    0.65 &    1.1  &    7.40 &   92.81 & [$SiO_4$]       &        65.84 \\
    3 &  As2S3         &    3.20 &    2.70 &    1.46 &     0.26 &    0.17 &    2.0 &    19.07 &   32.10  & [$S$]            &     246.03 \\
    4 &  LASF             &    5.79 &    5.64 &    3.60 &    1.46 &    0.92 &    0.4 &    12.65 &167.95   & [$LASF$]      &     221.30 \\
    5 &  SF4               &    4.78 &    3.78 &    2.24 &     0.72 &    0.48 &    1.1 &    8.40  &  136.17 & [$Si_2O_5$] &    116.78 \\
    6 &  SF59            &    6.26 &    3.32 &    1.92 &     0.77 &    0.49 &    1.0 &    14.05  &   92.81 & [$SiO_4$]  &     158.34 \\
    7 &  V52               &    4.80 &    4.15 &    2.25 &     0.87 &    0.52 &    1.7 &    8.37   &  167.21 & [$ZrF_4$] &     182.28 \\
    8 &  BALNA         &    4.28 &    4.30 &    2.30 &    0.75 &    0.45 &    2.1 &        6.87  &  167.21 & [$ZrF_4$]  &      140.79 \\
    9 &  LAT               &    5.25 &    4.78 &    2.80 &    1.13 &    0.65 &    1.4 &     9.16    &  205.21 & [$ZrF_6$] &     215.69 \\
   10 &  a-Se             &    4.30 &    2.00 &    1.05 &    0.25 &    0.14  &    2.0 &   18.23 &   78.96 & [$Se$]          &     78.96 \\
   11 &  Se75Ge25  &    4.35 &    0.00 &    1.24 &               &    0.15 &    1.0 &    22.19 &  77.38  & [$Se_3Ge_1$] &     77.38 \\
   12 &  Se60Ge40    &    4.25 &    2.40*&  1.44* &            &    0.16 &    0.4 &      23.56 &76.43 & [$Se_3Ge_1$] &       76.43 \\
   13 &  LiCl:7H2O    &    1.20 &    4.00 &  2.00* &      0.62 &    0.39 &    1.4 &    4.75  &  131.32 & [$Li(H_2O)Cl_3$] &168.50 \\
   14 &  Zn-Glass      &    4.24 &    4.60 &    2.30 &    0.70 &    0.38 &    2.2 &      7.71   &  103.41 & [$ZnF_2$] &    103.41 \\
   15 &  PMMA          &    1.18 &    3.15 &    1.57 &      0.39 &    0.27 &    0.6 &      6.10 &  102.78 & [$PMMA$]  &    102.78 \\
   16 &  PS                &    1.05 &    2.80 &    1.50 &    0.20 &     0.13  &    2.8 &    6.03 &   27.00  & [$CH-CH2$] &    105.15 \\
   17 &  PC                &    1.20 &    2.97 &   1.37* & 0.28 &    0.18  &    0.9 &      6.00    & 77.10  & [$C_6H_5$]  &     252.24 \\
   18 &  ET1000        &    1.20 &    3.25 &              &    0.35 &    0.22 &    1.1 &      4.91 &   77.10   & [$C_6H_5$]   &  77.10 \\
\hline
\end{tabular}
\end{ruledtabular}
\end{center}
\end{sidewaystable}

\newpage

\begin{table}[ht!]
\caption{\label{tab:table II} {\bf Comparison of various molecular and block length Scales for $M=M_1$:} The table displays the values for $R_m$ (eq.(\ref{ome0}), $R_0$ given by eq.(\ref{r05}) and block-ratio $R_0/R_v$ using $R_v \approx R_m$ and the block-radius $t$. Here $R_{0,a}$ and $t_a$ correspond to $R_0$ and $t$ from  longitudinal and transverse speeds of sound, with $a \equiv l, t$. These results are also illustrated in figure 1(a) and 1(b) separately for $R_{0,l}$ and $R_{0,t}$.}
  \begin{center}
\begin{ruledtabular}
\begin{tabular}{lcccccccr}
Index & Glass  & $R_m$ & $R_{0,l}$ & $R_{0,t}$ & $R_{0,l}/R_v$ & $R_{0,t}/R_v$ & $t_{l}$ & $t_t$\\
Units   &   & $angs.$ & $angs.$ & $angs.$  & & &  $angs.$ & $angs.
$   \\
\hline
1 &  a-SiO2           &    2.79 &   10.39 &   10.72 &    3.73 &    3.85 &   10.03 &   10.52\\
 2 &  BK7             &    2.45 &    9.74 &    9.12 &    3.98 &    3.72 &    9.72 &    8.80\\
 3 &  As2S3         &    1.58 &    8.17 &    7.20 &    5.16 &    4.54 &    9.27 &    7.67\\
 4 &  LASF7          &    2.26 &    9.87 &    9.95 &    4.37 &    4.41 &   10.32 &   10.45\\
 5 &  SF4              &    2.24 &    9.22 &    8.53 &    4.11 &    3.80 &    9.35 &    8.31\\
 6 &  SF59            &    1.80 &    6.64 &    6.23 &    3.68 &    3.45 &    6.37 &    5.79\\
 7 &  V52             &    2.40 &    9.89 &    9.27 &    4.12 &    3.86 &   10.04 &    9.11\\
 8 &  BALNA         &    2.49 &   10.88 &   10.08 &    4.36 &    4.04 &   11.36 &   10.13\\
 9 &  LAT             &    2.49 &   10.47 &   10.60 &    4.20 &    4.25 &   10.72 &   10.92\\
10 &  a-Se            &    1.94 &   10.26 &    9.83 &    5.30 &    5.07 &   11.81 &   11.07\\
11 &  Se75Ge25   &    1.92 &     NaN &   12.69 &     NaN &    6.62 &     NaN &   16.32\\
12 &  Se60Ge40   &    1.92 &    8.43 &   13.82 &    4.38 &    7.18 &    8.82 &   18.51\\
13 &  LiCl:7H2O     &    3.51 &   13.53 &   11.61 &    3.85 &    3.31 &   13.28 &   10.56\\
14 &  Zn-Glass       &    2.13 &    9.02 &    8.54 &    4.23 &    4.01 &    9.28 &    8.54\\
15 &  PMMA          &    3.26 &   14.59 &   11.72 &    4.48 &    3.60 &   15.45 &   11.12\\
16 &  PS               &    2.17 &    8.95 &    7.87 &    4.13 &    3.63 &    9.09 &    7.49\\
17 &  PC               &    2.94 &   14.29 &   11.45 &    4.86 &    3.89 &   15.75 &   11.30\\
18 &  ET1000       &    2.94 &   12.23 &    0.00 &    4.16 &    0.00 &   12.47 &    0.00\\
  \hline
\end{tabular}
\end{ruledtabular}
\end{center}
\end{table}

\begin{table}[ht!]
\caption{\label{tab:table III} {\bf Various molecular and block length Scales for $M_2$:} The other details are same as in table II.  The results are also illustrated in figure 2(a) and 2(b) separately for $R_{0,l}$ and $R_{0,t}$.}
  \begin{center}
\begin{ruledtabular}
\begin{tabular}{lcccccccr}
Index & Glass  &  $R_m$ & $R_{0,l}$ & $R_{0,t}$ & $R_{0,l}/R_v$ & $R_{0,t}/R_v$ & $t_{l}$ & $t_t$ \\
Units   &    & $angs.$ & $angs.$ & $angs.$ &   &  & $angs.$ & $angs.$ \\
\hline
 1 &  a-SiO2          &    2.21 &    6.55 &    6.76 &    2.96 &    3.05 &    5.63 &    5.91\\
 2 &  BK7             &    2.18 &    7.75 &    7.25 &    3.55 &    3.32 &    7.30 &    6.61\\
 3 &  As2S3         &    3.12 &   31.75 &   27.98 &   10.17 &    8.96 &   50.62 &   41.87\\
 4 &  LASF7         &    2.47 &   11.86 &   11.96 &    4.79 &    4.83 &   12.98 &   13.15\\
 5 &  SF4             &    2.13 &    8.32 &    7.70 &    3.91 &    3.61 &    8.23 &    7.31\\
 6 &  SF59           &    2.16 &    9.48 &    8.90 &    4.40 &    4.13 &    9.94 &    9.04\\
 7 &  V52             &    2.47 &   10.48 &    9.82 &    4.24 &    3.98 &   10.79 &    9.79\\
 8 &  BALNA         &    2.35 &    9.70 &    8.99 &    4.12 &    3.82 &    9.85 &    8.78\\
 9 &  LAT             &    2.53 &   10.82 &   10.95 &    4.27 &    4.32 &   11.18 &   11.38\\
10 &  a-Se            &    1.94 &   10.26 &    9.83 &    5.30 &    5.07 &   11.81 &   11.07\\
11 &  Se75Ge25   &    1.92 &     NaN &   12.69 &     NaN &    6.62 &     NaN &   16.32\\
12 &  Se60Ge40   &    1.92 &    8.43 &   13.82 &    4.38 &    7.18 &    8.82 &   18.51\\
13 &  LiCl:7H2O     &    3.82 &   15.98 &   13.71 &    4.19 &    3.59 &   16.35 &   12.99\\
14 &  Zn-Glass       &    2.32 &   10.66 &   10.10 &    4.60 &    4.36 &   11.44 &   10.54\\
15 &  PMMA           &    3.26 &   14.59 &   11.72 &    4.48 &    3.60 &   15.45 &   11.12\\
16 &  PS                &    3.41 &   22.15 &   19.47 &    6.49 &    5.71 &   28.23 &   23.26\\
17 &  PC                &    4.37 &   31.49 &   25.24 &    7.21 &    5.78 &   42.28 &   30.34\\
18 &  ET1000        &    2.94 &   12.23 &    0.00 &    4.16 &    0.00 &   12.47 &    0.00\\
\hline
\end{tabular}
\end{ruledtabular}
\end{center}
\end{table}

\newpage

\begin{table}[ht!]
\caption{\label{tab:table IV} Comparison of  theoretical and experimental  $\gamma_{m} \over c$-values for 18 glasses:
Here the columns $3, 4$ display the values obtained from eq.(\ref{r006}) for $M_1, M_2$ respectively,  with $A_H$ given in table I, (all values in units of ${ev\over km/sec}$). The colums $5, 8$ display the values for ${\gamma_l \over c_l}$ and ${\gamma_t \over c_t}$  with $c_l, c_t, \gamma_l, \gamma_t$ data taken from \cite{mb}.  The  ${\gamma_l \over c_l}$ and ${\gamma_t \over c_t}$-values shown in colums $ 6, 7, 9, 10$  are obtained from eq.(\ref{ts}) with $C_l, C_t$ values taken from \cite{plt}  and $\overline{P}$ given in $column 8$ of table I (same as in \cite{mb}).}
\begin{center}
\begin{ruledtabular}
\begin{tabular}{lccccccccr}
Index & Glass & ${\gamma\over c}$ & ${\gamma\over c}$ & ${\gamma_l\over c_l}$ & ${\gamma_l\over c_l}$ & ${\gamma_l\over c_l}$ & ${\gamma_t \over c_t}$ & ${\gamma_t\over c_t}$ & ${\gamma_t\over c_t}$ \\
& & $M_1$, eq.(\ref{r006}) & $M_2$, eq.(\ref{r006}) & \cite{mb} &   \cite{plt} & \cite{plt} &  \cite{mb} & \cite{plt} & \cite{plt}\\
\hline
1 &  a-SiO2          &    0.16 &       0.11 &    0.18 &    0.18 &    0.17 &    0.17 &    0.17 &    0.18 \\
 2 &  BK7              &    0.15 &       0.13 &    0.15 &    			&    		 &    0.17 &           &    \\
 3 &  As2S3           &    0.14 &      0.39 &    0.10 &    0.12  &    0.09 &    0.12 &            &     \\
 4 &  LASF             &    0.30 &      0.34 &    0.26 &    0.33 &    		&    0.26 &             &    \\
 5 &  SF4               &    0.20 &      0.18 &    0.19 &    0.00 &    		&    0.21 &    			&    \\
 6 &  SF59             &    0.20 &      0.27 &    0.23 &    0.00 &    		&    0.26 &    			&    \\
 7 &  V52               &    0.22 &     0.23 &    0.21 &    0.26 &    		 &    0.23 &    0.24  &   \\
 8 &  BALNA           &    0.20 &     0.18 &    0.17 &    0.00 &    		 &    0.20 &     		&    \\
 9 &  LAT               &    0.25 &     0.26 &    0.24 &    0.00 &    		&    0.23 &    			&    \\
10 &  a-Se             &    0.19 &     0.19 &    0.12 &    0.17 &   			 &    0.13 &    0.15 &    \\
11 &  Se75Ge25     &    0.26 &    0.26 &            &            &            &    0.12 &    			&    \\
12 &  Se60Ge40     &    0.27 &    0.27 &    0.23 &            &           &    0.11 &    			 &    \\
13 &  LiCl:7H2O     &    0.15 &   0.17 &    0.16 &    0.15 &    		   &    0.19 &             	&     \\
14 &  Zn-Glass      &    0.17 &   0.19 &    0.15 &    0.00 &    		   &    0.17 &      		 &    \\
15 &  PMMA           &    0.15 &    0.15 &    0.12 &    0.17 &    0.15 &    0.17 &    0.20 &    0.21\\
16 &  PS                 &    0.07 &    0.15 &    0.07 &    0.00 &    0.11 &    0.09 &            &     \\
17 &  PC                &    0.13 &    0.23 &    0.09 &    0.13 &    			&    0.13 &            &     \\
18 &  ET1000         &    0.11 &    0.11 &    0.11 &    0.14 &            &            &            &     \\
\hline
\end{tabular}
\end{ruledtabular}
\end{center}
\end{table}

\newpage

\begin{table}[ht!]
\caption{\label{tab:table V} {\bf Theoretically predicted boson frequency  for $M=M_1$:}
Theoretically $t_a=R_{0a}$ with $a \equiv l, t$ and therefore corresponding $\omega_d$ 
values are also expected to be same. However as table II indicates, $t_l$ can differ from $t_t$ for some cases. This motivates us to display here  the values for $\omega_{ir}$ (eq.(\ref{omd}), all values in $cm^{-1}$, labelled as $\omega_{ir,l1}, \omega_{ir,l2}, \omega_{ir,t1}, \omega_{ir,t2}$ for four possible mean free path $l \sim 2 R_{0l}, 2 t_l, R_{0t}, 2 t_t$, respectively, with $R_{0,a}$ and $t_a$ given in table II. A comparison of the data indicates $\omega_{ir,l1}=\omega_{ir,l2}$ and $\omega_{ir,t1}= \omega_{ir,t2}$; this is indeed expected based on our theoretical prediction $R_{0l}=t_l$ and $R_{0t}=t_t$.  
The data also indicates the closeness of $\omega_{ir,t}$, obtained from tranverse data,  to their experimental counterparts for boson peak frequency $\omega_{bp,exp}$ (available only for few glasses). The last colum displays the values for $e_0/\hbar$ theoretically conjectured to be of the same order as $\omega_{bp}$. }
  \begin{center}
\begin{ruledtabular}
\begin{tabular}{lccccccccr}
Index & Glass  & $\omega_{ir,l} $ & $\omega_{ir,l} $ & $\omega_{ir,t} $ & $\omega_{ir,t} $ & $\omega_{bp,exp} $   & $e_0/\hbar$ \\
m $l=$   &   &  $R_{0l}$ & $t_l$ &  $R_{0t}$  & $t_t$  &  \\
\hline
 1 &  a-SiO2          &   93.02 &   96.34 &   59.06 &   60.21 &  52 \cite{stol}        & 52.55     \\
 2 &  BK7             &  106.08 &  106.34 &   69.48 &   72.00 &              & 61.65          \\
 3 &  As2S3           &   55.09 &   48.52 &   33.81 &   31.73 &  26 \cite{nema}         & 149.87    \\
 4 &  LASF7           &   95.25 &   91.10 &   60.28 &   57.41  &  80 \cite{prat}        & 126.24    \\
 5 &  SF4             &   68.31 &   67.39 &   43.79 &   44.92    &             & 70.05       \\
 6 &  SF59            &   83.32 &   86.87 &   51.36 &   55.28   &            & 109.33   \\
 7 &  V52             &   69.92 &   68.87 &   40.45 &   41.16   &             &69.74     \\
 8 &  BALNA           &   65.87 &   63.06 &   38.04 &  37.84  &            & 57.26     \\
 9 &  LAT             &   76.11 &   74.28 &   44.04 &    42.73   &           & 76.30      \\
10 &  a-Se            &   32.48 &   28.23 &   17.80 &   15.81   & 18 \cite{novi}       & 111.19   \\
11 &  Se75Ge25        &         &              &   16.29 &  12.66   &           & 207.36   \\
12 &  Se60Ge40        &   47.47 & 45.37 &   17.37 &  12.96   &             & 226.28    \\
13 &  LiCl:7H2O       &   49.26 &   50.20 &   28.70 &   31.57 &  60  \cite{kgb}      & 39.57     \\
14 &  Zn-Glass         &  85.02 &   82.65 &   44.91 &   44.87  &.           & 64.28      \\
15 &  PMMA            &   35.98 &   33.99 &   22.32 &   23.53  & 15-20\cite{ryz}  & 50.81   \\
16 &  PS              &   52.15 &   51.33 &   31.78      &   33.37   &   17 \cite{pala}      & 50.29    \\
17 &  PC              &   34.64 &   31.44 &   19.94 &   20.21      &              & 49.99      \\
18 &  ET1000      &   44.27 &   43.42  &  &             &              &  40.94 \\
\end{tabular}
\end{ruledtabular}
\end{center}
\end{table}

\begin{figure}[ht!]
\vspace{-3in}
	\centering
	\includegraphics[width=1.2\textwidth,height=1.7\textwidth]{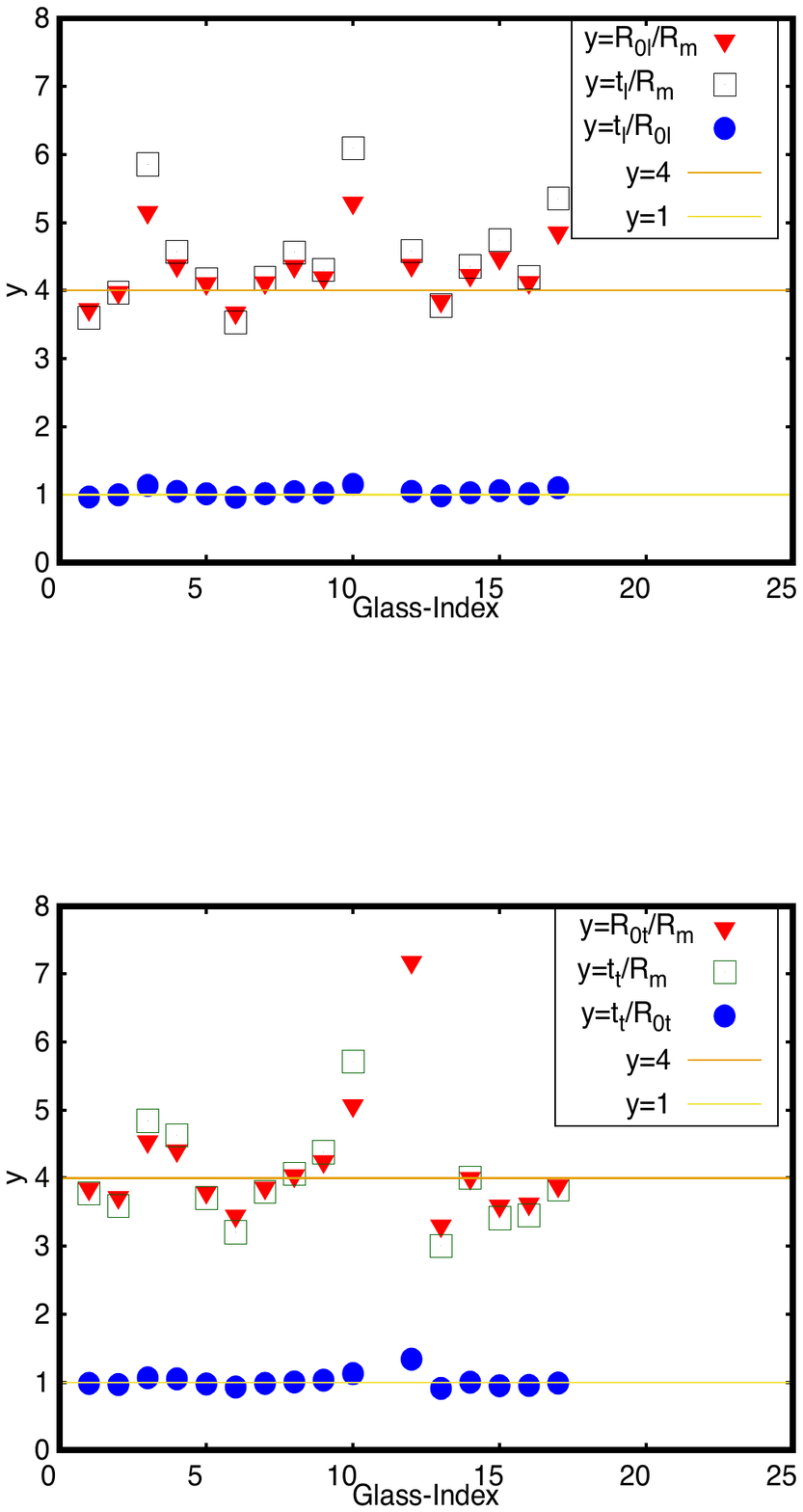}
	\caption{{\bf  Ratios ${R_{0,a}\over R_m}$ and ${t_a\over R_{0,a}}$ with $M=M_1$ and ($a=l,t$),  for $18$ glasses}: Here the solid lines $y=1$ and $y=4$ are shown for visual guidance only.  Clearly the collapse of all $18$ data-points on $y=1$ confirms the theoretical prediction $R_{0,a}=t_a$. Further, as the display around line $y=4$ indicates, the prediction corresponding to ${R_{0,a}\over R_v}=4$ with $R_v \approx R_m$ is satisfied by most glasses except for a few of them.   } 
\label{fig1}
\end{figure}

\begin{figure}[ht!]
\vspace{-3in}
	\centering
	\includegraphics[width=1.2\textwidth,height=1.7\textwidth]{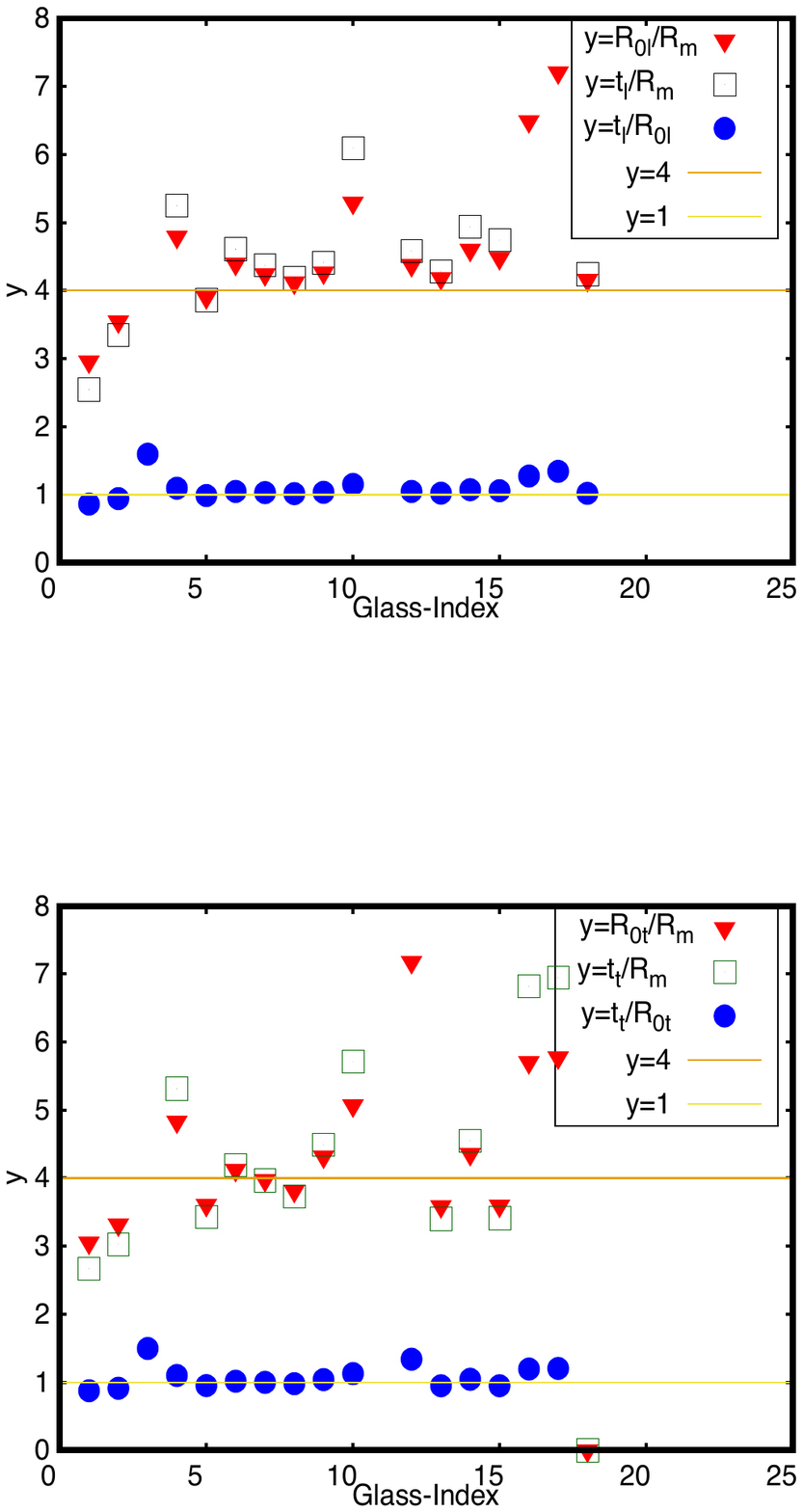}
	\caption{{\bf  Ratios ${R_{0,a}\over R_m}$ and ${t_a\over R_{0,a}}$ with $M=M_2$ and ($a=l,t$),  for $18$ glasses}: the details here are same as in figure 1 .} 
\label{fig2}
\end{figure}

\begin{figure}[ht!]
	\centering
	\includegraphics[width=0.8\textwidth,height=1.\textwidth]{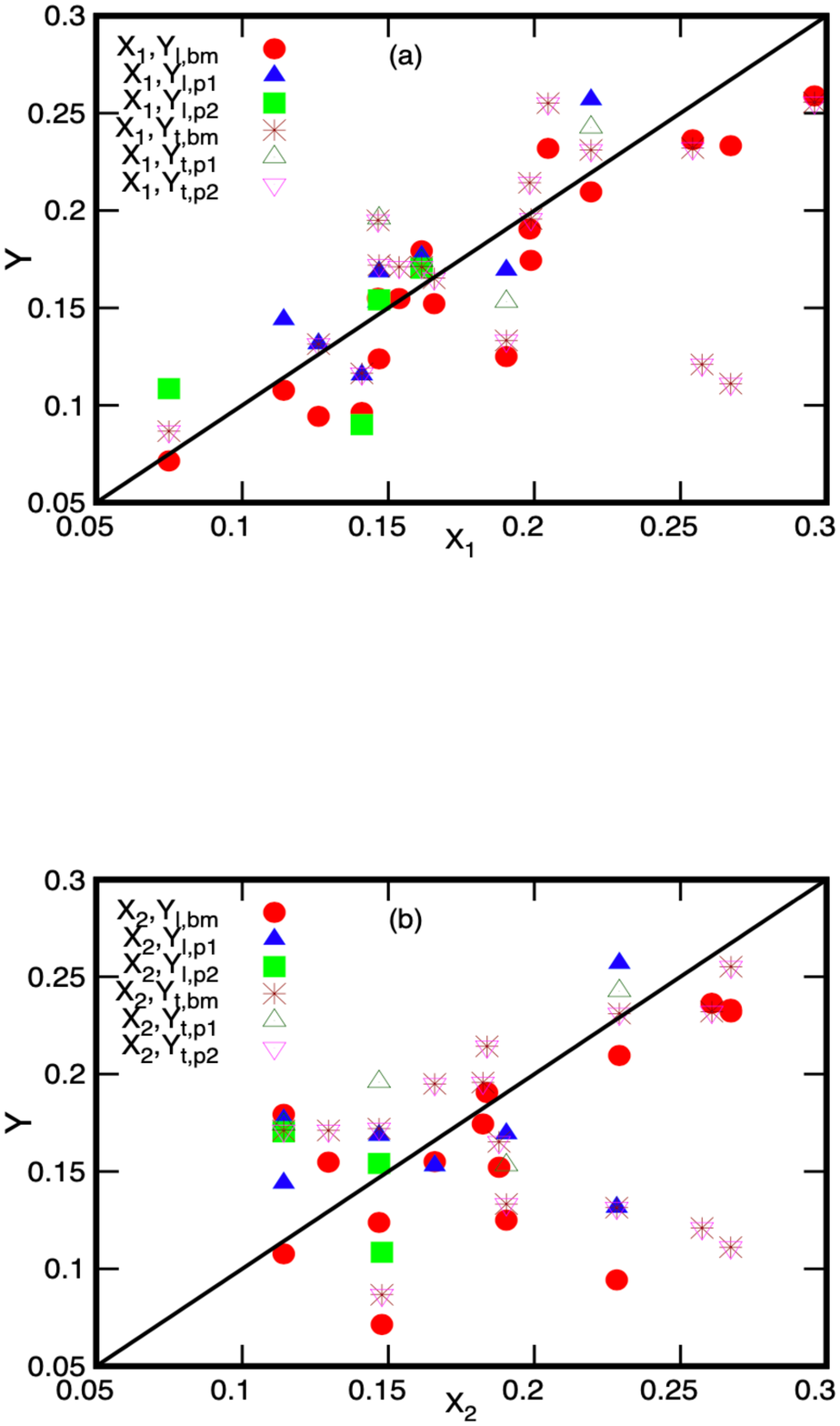}
	\caption{{\bf  Comparison of  theoretically predicted ${\gamma_m \over c}$ with experimnetal data}: 
here the  ${\gamma_m \over c}$-values from eq.(\ref{r006}) correspond to $x$-coordinates (labelled as $X_n$ with $n=1$ for $M=M_1$ and $n=2$ for $M=M_2$). The $y$ coordinates of each point correspond to experimental data, labelled as $Y_{a,xx} ={\gamma_a \over c_a}$ with ${\gamma_a\over c_a}$, taken from \cite{mb} (for $xx=bm$) and from \cite{plt} (accoustic data for  $xx=p1$ and flexural data for  $xx=p2$).}
\label{fig3}
\end{figure}



\begin{thebibliography}{10}



\bibitem{plt}
R.O.Pohl, X.Liu and E.Thompson, Rev. Mod. Phys. 74, 991, (2002). 


\bibitem{mb}
J.F. Berret and M. Meissner, Z. Phys. B-Condensed Matter 70, 65, (1988).


\bibitem{and}
P.W. Anderson, B.I. Halperin and C.M. Verma, Philos. Mag. 25, 1, (1972). 

\bibitem{phil}
W.A. Phillips, Two Level States in Glass, rep. Prog. Phys. 50, 1657, (1987);
R. Hunklingers and A.K.Raychadhuri, in Progr. Low-Temp. Phys. (ed. D. F. Brewer, Elsevier, Amsterdam), 9, 265, r1986; J. Jackle ,  {\it Amorphous Solids: Low-Temperature Properties}, (Springer, Berlin) 1981.


\bibitem{spm}
V.G.Karpov, M.I.Klinger, F.N.Ignatiev, Sov. Phys. JETP 57, 439, (1983).



\bibitem{buch}
U. Bucheanau, Y. M. Galperin, V. Gurevich, D. Parashin, M. Ramos and H. Schober, Phys. Rev. B 46, 2798, (1992); 
43, 5039, (1991)


\bibitem{paras}
D. A. Parashin, Phys. Rev. B, 49, 9400, (1994). 


\bibitem{gure}
V. Gurevich, D. Parashin and H. Schrober, Phys. Rev. B, 67, 094203, (2003). 



\bibitem{frac}
 S. Alexander and R. Orbach, J. Phys. (Paris) Lett. L625, 34, (1982). 

\bibitem{gg1}
 J.E.Graebner, B.Golding and L.C.Allen, Phys. Rev. B, 34, 5696, (1986).
 

\bibitem{sdg}
W. Schirmacher, G. Diezemann and C. Ganter, Phys. Rev. Lett. 81, 136, (1998).

\bibitem{schi}
W. Schirmacher, Europhys. Lett. 73, 892, (2006).

\bibitem{emt}
M. Wyart, Euro. Phys. Lett., 89, 64001, (2010). 

\bibitem{degi}
E. DeGiuli, A. Laversanne-Finot, G. During, E. Lerner and M. Wyart, Soft Matter, 10, 5628, (2014).


\bibitem{grig}
T. Grigera, V. Martin-Mayor, G. Parisi and P. Verrocchio, Nature, 422, 289, (2003). 


\bibitem{du}
E. Duval, A. Boukenter and T. Achibat, J. Phys. Condensed Matter, 2, 10227, (1990). 

\bibitem{vdos}
V. K. Malinovsky, V. N. Novikov, P.P. Parashin, A.P. Solokov and M.G. Zemlyanov, Europhys. Lett., 11, 43 (1990). 


\bibitem{ell3}
S.R.Elliott, Europhys. Lett. 19, 201 (1992).


\bibitem{sksq}
A. P. Sokolov, A. Kisliuk, M. Soltwisch, and D. Quitmann Phys. Rev. Lett. 69, 1540 (1992).


\bibitem{vl}
D. Vural and A.J.Leggett, J. Non crystalline solids, 357, 19, 3528, (2011). 



\bibitem{dl}
Z. Dee and A. J. Leggett, arXiv: 1510:05528v1.

\bibitem{lg1}
A. J. Leggett and D. Vural, J. Phys. Chem. B., 42,117, (2013).



\bibitem{lern}
E. Lerner, G. Düring, and E. Bouchbinder, Phys. Rev. Lett., 117, 035501, (2016).

\bibitem{mizu}
H. Mizuno, H. Shiba and A. Ikeda, PNAS 114, E9767, (2017). 


\bibitem{mg}
G. Monaco and V. M. Giordano, PNAS.0808965106.

\bibitem{bb1}
P. Shukla, arXiv:2008.12960.

\bibitem{qc1}
P. Shukla, arXiv:2009:00556

\bibitem{ell}
S.R.Elliott, Nature, 354, 445, (1991); A. Uhlhert and S.R. Elliott, J. Phys.: Condens. Matter 6, L99, (1994).

\bibitem{blhf} C. Buchner, L. Lichenstein, H.-J. Freund, {\it noncontact atomic force microscopy, nanoscience technology, S. Moira et. al. (eds.)}, Springer int. Pub. 2015, DOI. $10.11007/978-3-319-15588-3_16$.

\bibitem{ajs}
A.J.Stone, {\it The theory of intermolecular forces}, Oxford scholarship online, Oxford university Press, U.K. 2015.

\bibitem{ell2}
A. Uhlhert and S.R. Elliott, J. Phys.: Condens. Matter 6, L99, (1994).






\bibitem{isra}
J. Israelachvili, Chapter 11, {\it Intermolecular and Surface Forces}, 3rd ed. Academic Press, (2011).


\bibitem{fr2000}
R.H. French, J. Am. Ceram. Soc., 83, 2117, (2000).

\bibitem{ham}
H.C. Hamaker, Physica IV, no 10, 1058, (1937).


\bibitem{chum}
A. I. Chumakov et al., Phys. Rev. Lett. 106, 225501 (2011)

\bibitem{tara}
 S. N. Taraskin, Y. L. Loh, G. Natarajan, and S. R. Elliott, Phys. Rev. Lett. 86, 1255 (2001).
 



\bibitem{wang}
Y. Wang, L. Hong, Y. Wang, W. Schirmacher and J. Zhang, Phys. Rev. B 98, 174207 (2018). 
 
\bibitem{nie}
Y. Nie, H. Tong, J. Liu, M. Zu, N. Xu, Front. Phys. 12, 126301 (2017)



\bibitem{pala}
D. A. Parashin and C. Laermans, Phys. Rev. B 63,  132203, (2001).

\bibitem{stol}
R.H. Stolen, Phys.Chem.Glasses, 11, 83,  (1970).

\bibitem{nema}
R.J. Nemanich, Phys.Rev.B, 16, 1665, (1977).



\bibitem{novi}
V.N. Novikov, and A.P. Sokolov, Sol.State Comm., 77, 243, (1991).



\bibitem{prat}
 J.L. Prat, F. Terki, and J. Pelous, Phys.Rev.Lett., 77, 755, (1996).

\bibitem{kgb}
K.G. Breitschwerdt, and S. Gut, Proc. 12 Intern. Conf. on acoustic, Toronto, G2-6, (1986).


\bibitem{ryz}
V A Ryzhov,  Phys Astron Int J. 3, 123, (2019).


\bibitem{allen}
P.B. Allen, J.L.Feldman, J.Fabian and F.Wooten, Phil. Mag. B 79, 1715 (1999).

\bibitem{bkp}
Y.M.Beltukov, V.I.Kozub and D. A. Parashin, Phys. Rev. B 87,  134203, (2013).

\bibitem{shin}
H. Shintani and Y. Tanaka, Nat. Mater. 7, 870 (2008). 

\bibitem{bfpt}
Y. M. Beltukov, C. Fusco, D. A. Parshin, and A. Tanguy,Phys. Rev. E 93, 023006 (2016).

\end{thebibliography}
\end{document}